\documentclass[12pt,a4paper,oneside,reqno]{amsart}

\usepackage{lmodern}
\usepackage[margin=2.5cm]{geometry}
\usepackage{setspace}
\usepackage{amssymb}
\usepackage{color}
\usepackage[normalem]{ulem} 
\usepackage{mathtools}
\usepackage{lineno}
\usepackage{hyperref}

\author{Gregory J. Galloway}
\address{Department of Mathematics, University of Miami, Coral Gables, FL, USA.}
\email{galloway@math.miami.edu}

\author{Abraão Mendes}
\address{Instituto de Matemática, Universidade Federal de Alagoas, Maceió, AL, Brazil.}
\email{abraao.mendes@im.ufal.br}

\title[Expanding horizons]{Aspects of the geometry and topology\\ of expanding horizons}

\newtheorem{thm}{Theorem}[section]
\newtheorem{prop}[thm]{Proposition}

\newtheorem{Def}[thm]{Definition}

\theoremstyle{remark}
\newtheorem{rmk}[thm]{Remark}

\DeclareMathOperator{\tr}{tr}
\DeclareMathOperator{\divergence}{div}
\newcommand\ric{{\rm Ric}}

\newcommand{\p}{\partial}

\renewcommand{\div}{\divergence}
\renewcommand\th{\theta}
\renewcommand\S{\Sigma}

\newcounter{mnotecount}

\newcommand{\bbR}{\mathbb{R}}
\newcommand{\N}{\mathcal{N}}

\begin{document}

\raggedbottom

\numberwithin{equation}{section}

\setstretch{1.2}

\begin{abstract}
The aim of this paper is to extend some basic results about marginally outer trapped surfaces to the context of surfaces having general null expansion. Motivated in part by recent work of Chai-Wan, we introduce the notion of $\mathfrak{g}$-stability for a general closed hypersurface $\Sigma$ in an ambient initial data set and prove that, under natural energy conditions, $\Sigma$ has positive Yamabe type, that is, $\Sigma$ admits a metric of positive scalar curvature, provided $\Sigma$ is $\mathfrak{g}$-stable. A similar result is obtained when $\Sigma$ is embedded in a null hypersurface of a spacetime satisfying the dominant energy condition. Area bounds under similar conditions are obtained in the case where $\S$ is $2$-dimensional. Conditions implying $\mathfrak{g}$-stability are also discussed. Finally, we obtain a spacetime positive mass theorem for initial data sets with compact boundary $\Sigma$ of positive null expansion, assuming that the dominant energy condition is sufficiently strict near $\Sigma$. This extends recent results of Galloway-Lee and Lee-Lesourd-Unger.
\end{abstract}

\maketitle

\section{Introduction}

Marginally outer trapped surfaces are objects of considerable interest at the interface of spacetime geometry and the physics of black holes. The notion of a marginally outer trapped surface (or MOTS for short) was introduced early on in the development of the theory of black holes, in connection with gravitational collapse; see e.g.\ \cite{HE}. MOTSs appeared in a more purely mathematical context in the work of R.~Schoen and S.-T.~Yau~\cite{SYpmt2} concerning the existence of solutions of Jang’s equation, in connection with their proof of the positivity of energy. MOTSs may be viewed as spacetime analogues of minimal surfaces in Riemannian geometry and, despite the absence of a variational characterization like that for minimal surfaces, satisfy a number of analogous properties. 

MOTSs arise in various situations. For example, under suitable conditions, cross-sections of the event horizon in stationary black hole spacetimes (such as Schwarzschild and Kerr) are MOTSs. This can be roughly understood in terms of Hawking's area theorem. For dynamically evolving black hole spacetimes, the null geodesic generators of the event horizon have nonnegative expansion towards the future. However, in the steady state (stationary) limit, this expansion goes to zero. In dynamical black hole spacetimes (such as the Vaidya spacetime with null dust source), MOTSs typically occur inside the event horizon (as the boundary of the trapped region within a spacelike slice). In fact, there are general results showing that under appropriate conditions, MOTSs cannot occur outside the event horizon. (See Section~\ref{sec:2.3} for a brief discussion.)

The aim of this note is to extend some basic results about MOTSs (which e.g.\ apply to stationary black hole spacetimes) to results about surfaces having general (nonzero) null expansion, which may be viewed as applicable to dynamical black hole spacetimes. Surfaces of prescribed null expansion have recently been considered in \cite{Chai2}, in which Riemannian band width estimates are extended to spacetime initial data sets. 

In the next section, we present in some detail the key notions of stability for surfaces in expanding horizons, which we use in order to prove some results about the topology and area of these surfaces. These results are presented in Sections~\ref{sec:3.1} and~\ref{sec:3.2}. In Section~\ref{sec:3.3}, we consider a version of the spacetime positive mass theorem that applies to expanding horizons.

\medskip
\paragraph{\bf{Acknowledgements.}} 
The authors would like to thank the referee for many valuable comments on the first version of the paper. The work of the first named author was partially supported by the Simons Foundation, under Award No. 850541. The work of the second author was partially supported by the Conselho Nacional de Desenvolvimento Científico e Tecnológico - CNPq, Brazil (Grants 309867/2023-1 and 405468/2021-0) and the Fundação de Amparo à Pesquisa do Estado de Alagoas - FAPEAL, Brazil (Process E:60030.0000002254/2022).

\section{Surfaces of general null expansion: stability properties}\label{prelim}

All manifolds in this paper are assumed to be smooth, orientable and connected, unless otherwise stated.

An {\it initial data set} $(M,g,K)$ consists of an $n$-dimensional manifold $M$, $n \ge 3,$ equipped with a Riemannian metric $g$ and a symmetric $(0,2)$-tensor $K$. The main physical example is when $(M,g,K)$ is an initial data set in a spacetime (time-oriented Lorentzian manifold) $(\bar{M}, \bar{g})$, i.e.\ $M$ is a spacelike hypersurface in $\bar{M}$, with induced metric $g$ and second fundamental form $K$.

The {\it local energy density} $\mu$ and the {\it local current density} $J$ of an initial data set $(M,g,K)$ are given by,
\begin{align*}
\mu=\frac{1}{2}(S-|K|^2+(\tr K)^2)\quad\mbox{and}\quad J=\div(K-(\tr K)g),
\end{align*}
where $S$ is the scalar curvature of $(M,g)$. When $(M,g,K)$ is a spacetime initial data set, these quantities are given by $\mu = G(u,u)$ and $J = G(u,\cdot)$, where $G$ is the Einstein tensor $G={\rm Ric}_{\bar M}-\frac12 R_{\bar M}\bar g$ of $(\bar{M},\bar{g})$. Moreover, the inequality,
\begin{align}\label{DEC}
\mu \ge |J|
\end{align}
on $M$ is a consequence of the {\it spacetime dominant energy condition} (spacetime DEC),
\begin{align*}
G(X,Y)\ge0
\end{align*}
for all future-pointing causal vectors $X,Y\in T\bar{M}$. The inequality \eqref{DEC} plays the role of the dominant energy condition for initial data sets.

Though not strictly necessary, for the purpose of introducing certain concepts, we shall assume that $(M,g,K)$ is an initial data set in a spacetime $(\bar{M}, \bar{g})$. (In fact, this can always be arranged, cf.\ \cite{AM2009}.)
Then the second fundamental form $K$ of $M$ is given by,
\begin{align*}
K(X,Y)=\bar{g}(\bar{\nabla}_Xu,Y)
\end{align*} 
for $X,Y\in TM$, where $u$ is the future-pointing timelike unit normal to $M$ and $\bar{\nabla}$ is the Levi-Civita connection of $\bar{g}$. The mean curvature $\tr K$ of $M$ in $(\bar{M},\bar{g})$ is denoted by $\tau$. 

Let $\Sigma$ be a closed embedded hypersurface in $M$ with unit normal $\nu$ in $M$. By convention, we refer to $\nu$ as {\it outward pointing}. The {\it null expansion scalars} $\theta^+$, $\theta^-$ of $\Sigma$ in $M$ with respect to $\nu$ are defined by, 
\begin{align*}
\theta^\pm=\div_\Sigma\ell^\pm,
\end{align*}
where $\ell^+$, $\ell^-$ are the future-pointing null normal fields $\ell^\pm=u\pm\nu$ along $\Sigma$. The {\it null second fundamental forms} $\chi^+$, $\chi^-$ of $\Sigma$ in $(\bar{M},\bar{g})$ with respect to $\nu$ are defined by,
\begin{align*}
\chi^\pm(X,Y)=\bar{g}(\bar{\nabla}_X\ell^\pm,Y)
\end{align*}
for $X,Y\in T\Sigma$. Note, in terms of initial data, $\chi^\pm=K|_{\Sigma}\pm A$, where $A$ is the second fundamental form of $\Sigma$ in $M$ with respect to $\nu$. Also, $\theta^\pm=\tr\chi^\pm$.

\subsection{Stability with respect to initial data sets}

We now introduce a notion of stability for surfaces $\Sigma$ of general null expansion $\theta^+ = h$, which extends in a straightforward way the usual notion of stability of marginally outer trapped surfaces ($\theta^+ = 0$) in spacelike hypersurfaces.\ (As noted in the introduction, a slightly different notion of stability for surfaces of prescribed null expansion was considered in \cite{Chai2}; see also~\cite{Chai1}.) In order to discuss the stability properties of such surfaces (not necessarily MOTSs), we introduce a notation that allows one to compare pointwise the null expansions of two different surfaces.

For a fixed $\epsilon_0>0$ sufficiently small, the map
\begin{align*}
\Psi:[0,\epsilon_0]\times\Sigma\to M,\quad\Psi(t,p)=\exp_p(t\nu(p)),
\end{align*}
is well defined. Given $u\in C^\infty(\Sigma)$ positive with $\|u\|_{C^0}\le\epsilon_0$, the map
\begin{align*}
\Psi^u:[0,1]\times\Sigma\to M,\quad\Psi^u(t,p)=\exp_p(tu(p)\nu(p)),
\end{align*}
is also well defined. We denote $\Psi^u(t,\Sigma)$ by $\Sigma_t^u$ and the null expansion scalars of $\Sigma_t^u$ by~$\theta_u^\pm(t)$.

\begin{Def}
We say that $\Sigma$ is a {\it $\mathfrak{g}$-stable surface} ($\mathfrak{g}$ for generalized) if
\begin{align*}
\frac{\p\theta_u^+}{\p t}\big|_{t=0}\ge0\mbox{ for some }u>0\mbox{ with }\|u\|_{C^0}\le\epsilon_0.
\end{align*}
\end{Def}

\noindent
Heuristically speaking, $\Sigma$ is $\mathfrak{g}$-stable if there is an outward geodesic normal variation such that the null expansion is ``infinitesimally nondecreasing''. For MOTS, stability has essentially the same meaning, and is described by saying that a stable MOTS is ``infinitesimally outermost''. 

It is well known that (see e.g.\ \cite{AMS2008, AM2010, DLee}),
\begin{align*}
\frac{\p\theta_u^+}{\p t}\big|_{t=0}=Lu,
\end{align*}
where $L:C^\infty(\Sigma)\to C^\infty(\Sigma)$ is the elliptic operator,
\begin{align}\label{eq.stability.operator}
Lu=-\Delta u+2\langle X,\nabla u\rangle+\left(Q-|X|^2+\div X-\frac{1}{2}h^2+h\tau\right)u,
\end{align}
where $h=\theta_u^+(0)$ is the null expansion of $\Sigma_0^u=\Sigma$, $\tau = {\rm tr} K$ is the mean curvature of $M$,
\begin{align*}
Q=\frac{1}{2}R_\gamma-(\mu +J(\nu))-\frac{1}{2}|\chi^+|^2,
\end{align*}
$\gamma=\langle\,,\,\rangle$ is the induced metric on $\Sigma$, $R_\gamma$ is the scalar curvature of $\gamma$, and $X$ is the vector field tangent to $\Sigma$ that is dual to the 1-form $K(\nu,\cdot)|_\Sigma$. Moreover, there is a real number~$\lambda$, called the {\it principal eigenvalue} of $L$, satisfying $L\phi=\lambda\phi$ for some positive eigenfunction $\phi\in C^\infty(\Sigma)$, such that $\lambda\le\mbox{Re}(\mu)$ for any other eigenvalue $\mu$ of $L$. Also, the eigenspace of $L$ associated with $\lambda$ has dimension 1. 
By arguments essentially as in \cite{AMS2008} (see also\linebreak\cite[Theorem~A.10]{DLee}), one has that $\Sigma$ is $\mathfrak{g}$-stable if, and only if, $\lambda\ge0$.

It is of interest to have conditions that imply $\mathfrak{g}$-stability. A basic criterion for $\mathfrak{g}$-stability is obtained by extending the notion of {\it (locally) weakly outermost} for MOTSs to surfaces of prescribed null expansion. 
We say that $\Sigma$ is $\mathfrak{g}$-locally weakly outermost if for some 
$\epsilon_0 > 0$ sufficiently small, and for every $u\in C^\infty(\Sigma)$ positive with $\|u\|_{C^0}\le\epsilon_0$, there is no $t \in [0,1]$ such that the inequality $\theta^+_u(t) < h$ holds pointwise with respect to 
$\Psi^u$, i.e.\ such that
\begin{align*}
\theta^+_u(t)(p_{t,u})<h(p)\quad\text{for all}\quad p\in\Sigma,
\end{align*}
where $p_{t,u} = \Psi^u(t,p)$.
If $\Sigma$ is $\mathfrak{g}$-locally weakly outermost, then it is necessarily $\mathfrak{g}$-stable. 
In fact, if $\Sigma$ is not $\mathfrak{g}$-stable, then $\lambda<0$. Let $\phi>0$ be a principal eigenfunction of $L$. Without loss of generality, we may assume that $\|\phi\|_{C^0}\le\epsilon_0$. Therefore, 
\begin{align*}
\frac{\p\theta_\phi^+}{\p t}\big|_{t=0}=L\phi=\lambda\phi<0.
\end{align*}
Thus, since $\Sigma$ is compact, $\theta_\phi^+(t)<\theta_\phi^+(0)=h$ for $t>0$ sufficiently small.

\smallskip
We mention a simple criterion for the $\mathfrak{g}$-locally weakly outermost condition.

\begin{prop}\label{criterion1} 
Let $\Sigma$ be a closed embedded hypersurface in an initial data set $(M,g,K)$ with null expansion $\theta^+ = h$. Suppose there exists a variation 
$\Psi^{\hat{u}}$ of $\S$, $\hat{u}>0$, $\|\hat{u}\|_{C^0}\le\hat{\epsilon}_0$, such that 
$\theta^+_{\hat{u}}(\hat{t}) \ge h$ pointwise for all $\hat{t}\in [0,1]$. Then 
$\Sigma$ is 
$\mathfrak{g}$-locally weakly outermost.
\end{prop}

\proof 
Let $\Psi^u$, $u>0$, $\|u\|_{C^0}\le \epsilon_0$, be any variation of $\S$. Choose $\epsilon_0$ sufficiently small so that $\Psi^u([0,1] \times \S) \subset\Psi^{\hat{u}}([0,1] \times \Sigma)$. Suppose, by contradiction, that $\theta^+_u(t)<h$ pointwise, for some $t\in[0,1]$. By the compactness of $\Sigma_t^u$, there exists $\hat{t}$ such that $\Sigma_t^u$ lies to the inside of $\Sigma_{\hat{t}}^{\hat{u}}$ and so that they meet tangentially at some point $q = \Psi^u(t,p)$. Restricting the size of $\S$ to a small neighborhood of $p$, we may assume there exists a constant $a$ such that $\th_u^+(t)<a<\th^+_{\hat{u}}(\hat{t})$. But then the maximum principle for null expansion (\cite[Prop.~2.4]{AM2009},\linebreak \cite[Prop.~3.1]{AshGal}) would require $\th^+_u(t) = \hat{\th}^+_{\hat{u}}(\hat{t})$, which is a contradiction.
\qed

\smallskip 
We mention several examples. Consider the CMC spheres $r = r_0$ between the horizon and photon sphere ($2m < r_0 < 3m$) in the totally geodesic ($K=0$) time slice $t=t_0$ of Schwarzschild spacetime. The mean curvature (and hence the null expansion) of these spheres increases as $r$ increases from $2m$ to $3m$. It then follows from the proposition that each such sphere is $\mathfrak{g}$-locally weakly outermost, and hence stable. The fact that they are spheres is consistent with Theorem~\ref{thm.1.1}.

Now consider the manifold $M = \bbR \times V$, with metric $g=dt^2+\cosh^2t\,g_\S$, where $(\S, g_\S)$ is a higher genus surface of constant Gaussian curvature $-1$. Then $M$ has constant curvature $-1$. Consider the initial data set $(M,g,K=g)$, which one easily verifies is a vacuum initial data set; in fact, $\mu = 0$ and $J =0$. One further checks that the null expansion $h_t$ of the surfaces $\S_t = \{t\} \times \S$ increases from $0$ to $4$ as $t$ increases from $-\infty$ to $+\infty$. It again follows from the proposition that each $\S_t$ is $\mathfrak{g}$-locally weakly outermost, and hence stable. That the $\S_t$'s do not admit metrics of positive scalar curvature is consistent with Theorem~\ref{thm.1.1}. In this example, $\tau = 3 > \frac12 h_t$, and so illustrates the relevance of condition~(i) in Theorem~\ref{thm.1.1}. If we take $(M,g,K=ag)$, for any $a>1$, we have $\mu=3(a^2-1)$ and $J=0$. Now, $h_t$ increases from $-2+2a$ to $2+2a$ as $t$ increases from $-\infty$ to $+\infty$. In this case, $h_t\tau\ge6a(a-1)>3(a^2-1)$. This example illustrates the relevance of condition (ii) in Theorem~\ref{thm.1.1}.

\subsection{Stability with respect to null hypersurfaces}

We now extend the concept of $\mathfrak{g}$-stability for surfaces in initial data sets (e.g.\ in spacelike hypersurfaces) to surfaces in null hypersurfaces. Let $\N$ be a null hypersurface in a spacetime $(\bar{M},\bar{g})$ and $\Sigma$ be a closed hypersurface in $\N$ that is spacelike in $(\bar{M},\bar{g})$. Fix a future-pointing null vector field $\ell^-$ along $\Sigma$ that is orthogonal to $\Sigma$ and tangent to $\N$. 

As before, for $\epsilon_0>0$ sufficiently small, the map
\begin{align*}
F:[0,\epsilon_0]\times\Sigma\to\N,\quad F(t,p)=\exp_p(-t\ell^-(p))
\end{align*}
is well defined. Now, extend $\ell^-$ to a neighborhood of $\Sigma$ in $\N$ by,
\begin{align*}
\ell^-=-\frac{\p F}{\p t} 
\end{align*}
and define $\ell^+$ as the future-pointing null vector field along $\S$ that is normal to $\Sigma$ and satisfies $\langle\ell^+,\ell^-\rangle=-1$. In this case, the {\it null expansion scalars} $\theta^+$, $\theta^-$ of $\Sigma$ in $\N$ are defined by,
\begin{align*}
\theta^\pm=\div_\Sigma\ell^\pm.
\end{align*}

Clearly, the map
\begin{align*}
F^u:[0,1]\times\Sigma\to\N,\quad F^u(t,p)=\exp_p(-tu(p)\ell^-(p)),
\end{align*}
is also well defined for $u\in C^\infty(\Sigma)$ positive with $\|u\|_{C^0}\le\epsilon_0$. 
Extend $\ell^+$ to be the future-pointing null vector field that is normal to $\Sigma_t^u := F^u(t,\Sigma)$ for each $t$ and satisfies 
$\langle\ell^+,\ell^-\rangle=-1$. Denote the null expansion scalars of $\Sigma_t^u$ in $\N$ by $\theta_u^\pm(t)=\div_{\Sigma_t^u}\ell^\pm$. In this case, from well-known formulas (\cite{AMS2008,AM2010}), we have,
\begin{align*}
\dfrac{\p\theta_u^+}{\p t}\big|_{t=0}=L_-u,
\end{align*}
where $L_-:C^\infty(\Sigma)\to C^\infty(\Sigma)$ is the elliptic operator,
\begin{align}\label{eq.stability.operator.null}
L_-u=-\Delta u+2\langle X,\nabla u\rangle+\left(\frac{1}{2}R_\gamma-G(\ell^+,\ell^-)-|X|^2+\div X+h\theta^-\right)u,
\end{align}
where $h=\theta_u^+(0)$ and $\theta^-=\theta_u^-(0)$ are the null expansion scalars of $\Sigma$ in $\N$.

\smallskip
Similarly to the initial data case, we make the following definition.
\begin{Def}
We say that $\Sigma$ is a {\it $\mathfrak{g}$-stable surface} in $\N$ if 
\begin{align*}
\frac{\p\theta_u^+}{\p t}\big|_{t=0}\ge0\mbox{ for some }u>0\mbox{ with }\|u\|_{C^0}\le\epsilon_0.
\end{align*}
\end{Def}

Again, as in the initial data case, one has that $\Sigma$ is $\mathfrak{g}$-stable if, and only if, 
$\lambda_1(L_-)\ge0$, where $\lambda_1(L_-)$ is the principal eigenvalue of $L_-$.\ A completely analogous definition of\linebreak $\mathfrak{g}$-locally weakly outermost holds in this null hypersurface case (just replace $\Psi^u$ by $F^u$), and again $\mathfrak{g}$-locally weakly outermost implies $\mathfrak{g}$-stability.

In a similar manner, we have the following null hypersurface version of Proposition~\ref{criterion1}.

\begin{prop}\label{criterion2}
 Let $\S$ be a closed embedded hypersurface in a null hypersurface $\N$. Suppose there exists a variation $F^{\hat{u}}$ of $\S$, $\hat{u}>0$, $\|\hat{u}\| _{C^0}\le\hat{\epsilon}_0$, such that $\th^+_{\hat{u}}(\hat{t}) \ge h$ pointwise for all $\hat{t}\in [0,1]$. Then $\S$ is 
$\mathfrak{g}$-locally weakly outermost.
\end{prop}

The proof is essentially the same as in Proposition~\ref{criterion1}, except that a variation of the maximum principle referenced there is needed, one that applies to null, rather than spacelike, hypersurfaces.\ That the maximum principle for null expansion holds in this case follows from \cite[Theorem~2]{Mars}, which is a consequence of the maximum principle for null hypersurfaces \cite{Gnullmp}. 

Vaidya spacetime, which may be viewed as a dynamical version of Schwarzschild spacetime, provides a nice illustration of Proposition~\ref{criterion2}. The metric in ingoing Eddington-Finkelstein coordinates $(v,r,\theta,\phi)$ is given by the line element,
\begin{align*}
ds^2 = -\left(1 - \frac{2M}{r}\right) dv^2 +2dvdr +r^2 d\Omega^2, 
\end{align*}
where $M = M(v)$. The dominant energy condition is satisfied if $M(v)$ is nondecreasing. 
Consider a null hypersurface $\N: v = v_0$. The $2$-spheres $\S_r$ in $\N$ have null expansion 
\begin{align*}
\th^+_r=\frac{r-2M}{r^2}
\end{align*}
with respect to the future directed outward null normal,
\begin{align*}
\ell^+=\frac{\p}{\p v}+\frac{1}{2}\left(1-\frac{2M}{r}\right)\frac{\p}{\p r},
\end{align*}
where $M = M(v_0)$ (see e.g.\ \cite{AK}). The null expansion $\th^+_r $ increases as $r$ increases from $2M$ to $4M$. Using the approximate location of the event horizon $\mathcal{H}$ (as analyzed in e.g.\ \cite{Nielsen}), it can be shown that, for suitable functions $M = M(v)$, the spherically symmetric intersection of $\mathcal{H}$ and $\N$ will lie well within this range. It then follows from Proposition~\ref{criterion2} that this intersection is $\mathfrak{g}$-locally weakly outermost, and hence stable.

\smallskip
The following will be important for the proofs of several of our results. It follows from a straightforward generalization of arguments in \cite{GS}; see \cite[Sect.~2]{MOTS06} and \cite[Sect.~2]{GalMen2018}.

\begin{thm}\label{lemma.main}
Let $(\Sigma,\gamma)$ be a closed Riemannian manifold and $L:C^\infty(\Sigma)\to C^\infty(\Sigma)$ be a differential operator of the form
\begin{align*}
Lu=-\Delta u+2\langle X,\nabla u\rangle+(\mathcal{Q}-|X|^2+\div X)u,
\end{align*}
where $\mathcal{Q}\in C^{\infty}(\S)$ and $X$ is a smooth tangent vector field on $\Sigma$. Suppose the principal eigenvalue of $L$ is nonnegative, $\lambda_1(L) \ge 0$. Then the following hold:
\begin{enumerate}
\item[(i)]
\begin{align}\label{eq.Rayleigh}
\int_{\S}(|\nabla \psi|^2 + \mathcal{Q} \psi^2)dA\ge 0
\end{align}
for all $\psi \in C^{\infty}(\S)$. Hence, the first eigenvalue $\lambda_1(L_0)$ of $L_0=-\Delta+\mathcal{Q}$ is nonnegative. Furthermore, if equality in \eqref{eq.Rayleigh} holds, and $\psi\not\equiv0$, then $\psi$ is an eigenfunction of $L_0$ associated with $\lambda_1(L_0)=0$;
\item[(ii)] If $\mathcal{Q}=\frac{1}{2}R_\gamma-\mathcal{P}$ for some function $\mathcal{P}\ge0$, then $\Sigma$ admits a metric of positive scalar curvature, unless $\lambda_1(L)=0$, $\mathcal{P}\equiv0$, and $(\Sigma,\gamma)$ is Ricci flat.
\end{enumerate}
\end{thm}

\subsection{Further comments on the $\mathfrak{g}$-stability condition}\label{sec:2.3}

As briefly noted in the introduction, under suitable conditions, cross sections $\S$ of the event horizon in stationary black hole spacetimes are MOTS (this follows from, for example, results in~\cite{ChruCosta}). Moreover, under general circumstances, such a MOTS $\S$ will be stable. For suppose not. Then $\S$ could be pushed outside the event horizon along a spacelike (or null) hypersurface to an outer trapped surface $\hat{\S}$. Now, it is a basic result in the theory of black holes that, for suitably defined black hole spacetimes satisfying the null energy condition, $\ric(X,X) \ge 0$ for all null vectors $X$ (in particular, satisfying the vacuum Einstein equations, $\ric\equiv 0$), trapped surfaces cannot exist outside the event horizon; see e.g.\ \cite[Prop.~12.2.2]{Wald}.\ The proof is easily modified to also rule out {\it outer} trapped surfaces that are homologous to the cross section $\S$, such as $\hat{\S}$.

The point we wish to make here is that essentially the same conclusion of stability can be reached for cross sections $\S$ of expanding horizons, provided there is sufficient matter in the vicinity of the black hole.

Let $\S$ be a cross section with null expansion $h$ bounded above by a constant $\th_0 >0$. If $\S$ is not stable, then it can be pushed outside the event horizon along a spacelike (or null) hypersurface to a surface $\hat{\S}$ having null expansion $\hat{h} < \th_0$. Suppose however that, due to the presence of matter, along each future complete outward null normal geodesic $s\longmapsto\eta(s)$ to $\hat{\S}$, with $\eta'(0)=\ell^+(p)$, $p=\eta(0)\in\hat{\S}$, 
\begin{align*}
\int_{0}^{\infty} \ric(\eta',\eta') ds \ge \th_0.
\end{align*}
Then the null expansion $\hat{h}$ of $\hat{\S}$ cannot satisfy $\hat{h} < \th_0$, for if it did then e.g.\ \cite[Prop.~2]{GalSen} would imply that there is a null focal point along every such $\eta$. However, arguing as in the proof of \cite[Prop.~12.2.2]{Wald}, there must exist at least one $\eta$ without a focal point.

\newpage

\section{Properties expanding horizons}

Here we present the results briefly mentioned in the introduction.

\subsection{The topology of expanding horizons}\label{sec:3.1}

\begin{thm}\label{thm.1.1}
Let $(M,g,K)$ be an initial data set and $\Sigma$ be a closed hypersurface in $M$. If $\Sigma$ is a $\mathfrak{g}$-stable surface in $(M,g,K)$ with null expansion 
$\theta^+=h \in C^{\infty}(\Sigma)$, $h\ge 0$, and either
\begin{enumerate}
\item[(i)] $\mu-|J|\ge0$ and $\tau \le \frac12 h$ along $\Sigma$, $h \not\equiv 0$,
or
\item[(ii)] $\mu-|J|\ge c_0$ and $h\tau\le c_0$ along $\Sigma$ for some constant $c_0>0$, 
\end{enumerate}
then $\Sigma$ admits a metric of positive scalar curvature.
\end{thm}

As is well known, there are many restrictions on the topology of manifolds that admit a metric of positive scalar curvature (see e.g.\ \cite{ChodoshLi} for a recent reference). We note that the condition 
on $\tau$ (which, recall, is the trace of $K$) in case~(i) is satisfied if $\tau \le 0$, and hence, in particular, if $M$ is {\it maximal}, i.e.\ $\tau = 0$. It is also satisfied, of course, if $h$ has a positive lower bound $h_0$ and $\tau \le \frac12 h_0$. A result related to Theorem~\ref{thm.1.1} was obtained in~\cite{Chai2} using a slightly different version of stability, one which is well-adapted to constructing surfaces of prescribed null expansion.

\proof
It follows from \eqref{eq.stability.operator} that 
\begin{align*}
Lu=-\Delta u+2\langle X,\nabla u\rangle+(\mathcal{Q}-|X|^2+\div X)u,
\end{align*}
where
\begin{align*}
\mathcal{Q}=Q-\frac{1}{2}h^2+h\tau=\frac{1}{2}R_\gamma-\mathcal{P}
\end{align*}
and
\begin{align}\label{eqn.P}
\mathcal{P}=\mu+J(\nu)+ \frac{1}{2}|\chi^+ |^2+\frac{1}{2}h^2 - h\tau.
\end{align}
\medskip
\noindent
\underline{Case (i)}: In this case, since $\mu+J(\nu) \ge \mu - |J| \ge 0$, we have
\begin{align*}
\mathcal{P}\ge\mu+J(\nu)+h\left(\frac{1}{2}h-\tau\right)\ge0
\end{align*} 
by the assumptions on $h$ and $\tau$. If $P\equiv 0$, then the above inequalities give that
\begin{align*}
h\left(\frac{1}{2}h - \tau \right) = -(\mu + J(\nu)).
\end{align*}
Substitution of this into \eqref{eqn.P} then implies $\chi^+ \equiv 0$, and hence $h=\tr\chi^+\equiv 0$, which is a contradiction. Thus, $P\not\equiv 0$.

\smallskip
\noindent
\underline{Case (ii)}: In this case, 
\begin{align*}
\mathcal{P} \ge \mu+J(\nu) - h\tau \ge 0,
\end{align*} 
since $\mu+J(\nu) \ge c_0$ and $h\tau \le c_0$. If $P \equiv 0$, then the above inequalities imply $h \tau = \mu+J(\nu)$. Substitution of this into \eqref{eqn.P} gives that $h \equiv 0$. On the other hand,
\begin{align*}
c_0 \ge h \tau = \mu+J(\nu) \ge c_0,
\end{align*} 
and hence $h \tau = c_0 > 0$, which is a contradiction. Thus, $P \not\equiv 0$. 

\smallskip
In either case, the result now follows from Theorem~\ref{lemma.main}.\qed

\medskip
The next result is a spacetime result, in which $\S$ is embedded in a null hypersurface. Physically, we may think of $\S$ as arising from the transverse intersection of the event horizon (assumed to be sufficiently smooth) with this hypersurface. The theorem assumes the (spacetime) dominant energy condition, and a natural condition on the inner null expansion of $\S$.

\begin{thm}\label{thm.1.2}
Let $(\bar{M},\bar{g})$ be a spacetime satisfying the dominant energy condition and $\N$ be a null hypersurface in $(\bar{M},\bar{g})$. If $\Sigma$ is a $\mathfrak{g}$-stable surface in $\N$ with null expansion scalars $\theta^+=h\ge0$, $h\not\equiv 0$, and $\theta^-<0$, then $\Sigma$ admits a metric of positive scalar curvature.
\end{thm}

\medskip
\proof Similarly to the proof of Theorem~\ref{thm.1.1}, it follows from \eqref{eq.stability.operator.null} that
\begin{align*}
L_-u=-\Delta u+2\langle X,\nabla u\rangle+\left(\mathcal{Q}-|X|^2+\div X\right)u,
\end{align*} 
where
\begin{align*}
\mathcal{Q}=\frac{1}{2}R_\gamma-\mathcal{P}
\end{align*}
and 
\begin{align*}
\mathcal{P}=G(\ell^+,\ell^-)-h\theta^-\ge-h\theta^- \ge 0.
\end{align*}
Above, we have used the dominant energy condition. Since $h\theta^-\not\equiv 0$, the result follows from Theorem~\ref{lemma.main}.\qed

\subsection{Area bounds for expanding horizons}\label{sec:3.2}

\begin{thm}
Let $(M,g,K)$ be a $3$-dimensional initial data set and $\Sigma$ be a closed surface in $M$. Suppose $\Sigma$ is a $\mathfrak{g}$-stable surface in $(M,g,K)$ with null expansion $\theta^+=h \in C^{\infty}(\Sigma)$, $h\ge 0$, such that $\tau \le \frac12 h$. If $\mu -|J| \ge c$ for some constant $c > 0$, then $\S$ is topologically a $2$-sphere and the area of $\S$ satisfies,
\begin{align}\label{ineqA}
A(\S) \le \frac{4\pi}{c}.
\end{align}
Furthermore, if $A(\Sigma)=4\pi/c$, then the following hold along $\Sigma$:
\begin{enumerate}
\item $\chi^+=0$ and, in particular, $\S$ is a MOTS;
\item $\Sigma$ is a round $2$-sphere of constant Gaussian curvature $\kappa_\S=c$;
\item $\mu+J(\nu)=\mu-|J|=c$ on $\Sigma$.
\end{enumerate}
\end{thm}

\begin{rmk}
Rigidity results deriving from equality in~\eqref{ineqA} have been examined in detail by the authors in~\cite{GalMen2018} and~\cite{GalMen2024}. We note, however, that equality is not realized for expanding horizons, $\theta^+ = h > 0$. 
\end{rmk}

\proof 
It follows from \eqref{eq.stability.operator} that 
\begin{align*}
Lu=-\Delta u+2\langle X,\nabla u\rangle+(\mathcal{Q}-|X|^2+\div X)u,
\end{align*}
where
\begin{align*}
\mathcal{Q}=\kappa_\Sigma-\mathcal{P}
\end{align*}
and
\begin{align*}
\mathcal{P} &=\mu+J(\nu)+\frac{1}{2}|\chi^+|^2+\frac{1}{2}h^2-h\tau\\
&\ge\mu-|J|+ h\left(\frac{1}{2}h-\tau\right)\ge c.
\end{align*} 

Hence, setting $\psi = 1$ in Theorem~\ref{lemma.main}~(i), we obtain,
\begin{align}\label{eqn.area}
c A(\S) \le \int_{\S}\mathcal{P}dA \le \int \kappa_{\S}\,dA = 4\pi (1-g(\Sigma)),
\end{align}
where $g(\Sigma)$ is the genus of $\Sigma$. These inequalities force $g(\Sigma)=0$, and the area bound follows.

If $A(\Sigma)=4\pi/c$, then all above inequalities must be equalities. In particular, 
\begin{itemize}
\item $\chi^+=0$;
\item $\mu+J(\nu)=\mu-|J|=c$ on $\Sigma$;
\item $\psi=1$ is an eigenfunction of $L_0=-\Delta+\mathcal{Q}$ associated with the first eigenvalue $\lambda_1(L_0)=0$, that is, 
\begin{align*}
\mathcal{Q}=\kappa_\S-\mathcal{P}=0\quad\therefore\quad\kappa_\S=\mathcal{P}=c.
\end{align*}
\end{itemize}
\qed

Similarly, we have:

\begin{thm}\label{thm:nullcase}
Let $(\bar{M},\bar{g})$ be a $(3+1)$-dimensional spacetime and $\N$ be a null hypersurface in $(\bar{M},\bar{g})$. Suppose $\Sigma$ is a $\mathfrak{g}$-stable surface in $\N$ with null expansion scalars $\theta^+=h\ge0$ and $\theta^-<0$. If $G(\ell^+,\ell^-) \ge c$ for some constant $c > 0$, then $\S$ is topologically a $2$-sphere and the area of $\S$ satisfies,
\begin{align*}
A(\S) \le \frac{4\pi}{c}.
\end{align*}
Furthermore, if  $A(\Sigma)=4\pi/c$, then the following hold along $\Sigma$:
\begin{enumerate}
\item $\S$ is a MOTS;
\item $\Sigma$ is a round $2$-sphere of constant Gaussian curvature $\kappa_\S=c$;
\item $G(\ell^+,\ell^-)=c$ on $\Sigma$.
\end{enumerate}
\end{thm}

\proof 
It follows from \eqref{eq.stability.operator.null} that
\begin{align*}
L_-u=-\Delta u+2\langle X,\nabla u\rangle+\left(\mathcal{Q}-|X|^2+\div X\right)u,
\end{align*} 
where
\begin{align*}
\mathcal{Q}=\kappa_\S-\mathcal{P}
\end{align*}
and 
\begin{align*}
\mathcal{P}=G(\ell^+,\ell^-)-h\theta^-\ge c.
\end{align*}
Inequalities \eqref{eqn.area} again apply.

If $A(\Sigma)=4\pi/c$, then $\mathcal{P}=G(\ell^+,\ell^-)=c$ and $-h\theta^-=0$ on $\Sigma$; in particular, $\Sigma$ is a MOTS, since $\theta^-<0$. Also, as before, $\mathcal{Q}=0$, that is, $\kappa_\Sigma=\mathcal{P}=c$.
\qed

\medskip
Theorem \ref{thm:nullcase} improves certain aspects of results in \cite{Hay}.

\subsection{A positive mass theorem for expanding horizons}\label{sec:3.3}

The next theorem pertains to the spacetime positive mass 
theorem~\cite{EHLS}. Recent results have extended this theorem to initial data sets with boundary \cite{GallowayLee,LeeLesourdUnger}, where the components are assumed to be weakly outer trapped ($\theta^+\le0$), or weakly inner untrapped ($\theta^{-}\ge0$). Here we show that the weakly outer trapped condition can be relaxed a bit, by requiring the dominant energy condition (for initial data sets) to be sufficiently strict near $\S$. 

\begin{thm}\label{PET}
Let $(M,g,K)$ be an n-dimensional, $3\le n\le 7$, complete asymptotically flat initial data set with compact boundary $\Sigma$. Assume that $(M,g,K)$ satisfies the DEC such that $\mu-|J|\ge c$ on a normal neighborhood $U\cong[0,2\epsilon]\times\Sigma$ of $\Sigma$ for some constant $c>0$. Suppose $\theta_0\coloneqq\sup_\Sigma\theta^+$ is positive, where $\theta^+$ is the null expansion of $\Sigma$ with respect to the normal pointing into $M$. Then $E\ge|P|$, provided $\theta_0\le c\,\epsilon/(2.18 + \epsilon\,\|\tau\|_{C^0(U)})$, where $(E,P)$ is the ADM energy-momentum vector of $(g,K)$.
\end{thm}

\proof On $U\cong[0,2\epsilon]\times\Sigma$, $g$ takes the form $g=dt^2+\gamma_t$, where $\gamma_t$ is the induced metric on $\Sigma_t\cong\{t\}\times\Sigma$.

Consider the modified initial data set $(M,g,\hat{K})$, where $\hat{K}=K-\frac{h}{n-1}g$. On $U$, the function $h$ is defined by,
\begin{align*}
h(t)=
\begin{cases}
\theta_0\exp\left(1-\dfrac{1}{1-(\frac{t}{\epsilon})^2}\right),& 0\le t<\epsilon,\\
0,& \epsilon\le t\le2\epsilon,
\end{cases}
\end{align*}
and $h\equiv0$ on $M\setminus U$. Clearly, $0\le h\le h(0)=\theta_0$. Furthermore, simple computations show that $h'(t)\le0$ and its minimum is attained at $t_0=3^{-\frac{1}{4}}\epsilon$ with $h'(t_0)\approx-2.17\,\theta_0/\epsilon$; in particular, $|Dh|\le 2.18\,\theta_0/\epsilon$.

Note that $\hat{\theta}^+\coloneqq\tr_\Sigma(\hat{K}+A)=\theta^+-\theta_0\le0$, where $A$ is the second fundamental form of $\Sigma$ in $(M,g)$ with respect to the normal pointing into $M$; that is, $\Sigma$ is weakly outer trapped in $(M,g,\hat{K})$. Further, the DEC clearly holds on $M\setminus U$. On $U$, a computation shows that (cf.\ \cite[Section~6]{LLU23})
\begin{align*}
\hat{\mu}-|\hat{J}|&=\mu+\frac{1}{2}\left(\frac{n}{n-1}h^2-2h\tau\right)-|J+Dh|\\
&\ge\mu-|J|+\frac{1}{2}\left(\frac{n}{n-1}h^2-2h\tau-2|Dh|\right)\\
&\ge c-\theta_0\left(\|\tau\|_{C^0(U)}+2.18\epsilon^{-1}\right)\ge0.
\end{align*}
The theorem then follows from \cite[Theorem~1.3]{LeeLesourdUnger}.

\newpage

\bibliographystyle{amsplain}
\bibliography{bibliography.bib}

\providecommand{\bysame}{\leavevmode\hbox to3em{\hrulefill}\thinspace}
\providecommand{\MR}{\relax\ifhmode\unskip\space\fi MR }
\providecommand{\MRhref}[2]{%
  \href{http://www.ams.org/mathscinet-getitem?mr=#1}{#2}
}
\providecommand{\href}[2]{#2}
\begin{thebibliography}{10}

\bibitem{AMS2008}
Lars Andersson, Marc Mars, and Walter Simon, \emph{Stability of marginally
  outer trapped surfaces and existence of marginally outer trapped tubes}, Adv.
  Theor. Math. Phys. \textbf{12} (2008), no.~4, 853--888. \MR{2420905}

\bibitem{AM2009}
Lars Andersson and Jan Metzger, \emph{The area of horizons and the trapped
  region}, Comm. Math. Phys. \textbf{290} (2009), no.~3, 941--972. \MR{2525646}

\bibitem{AM2010}
\bysame, \emph{Curvature estimates for stable marginally trapped surfaces}, J.
  Differential Geom. \textbf{84} (2010), no.~2, 231--265. \MR{2652461}

\bibitem{AshGal}
Abhay Ashtekar and Gregory~J. Galloway, \emph{Some uniqueness results for
  dynamical horizons}, Adv. Theor. Math. Phys. \textbf{9} (2005), no.~1, 1--30.
  \MR{2193368}

\bibitem{AK}
Abhay Ashtekar and Badri Krishnan, \emph{Dynamical horizons and their
  properties}, Phys. Rev. D (3) \textbf{68} (2003), no.~10, 104030, 25.
  \MR{2071054}

\bibitem{Chai1}
Xiaoxiang Chai, \emph{Hypersurfaces of prescribed null expansion}, 2021,
  arXiv:2107.12782.

\bibitem{Chai2}
Xiaoxiang Chai and Xueyuan Wan, \emph{Band width estimates of {CMC} initial
  data sets}, 2022, arXiv:2206.02624.

\bibitem{ChodoshLi}
Otis Chodosh and Chao Li, \emph{Recent results concerning topological
  obstructions to positive scalar curvature}, Perspectives in scalar curvature.
  {V}ol. 2, World Sci. Publ., Hackensack, NJ, [2023] \copyright 2023,
  pp.~215--230. \MR{4577915}

\bibitem{ChruCosta}
Piotr~T. Chru\'sciel and Jo\~ao~Lopes Costa, \emph{On uniqueness of stationary
  vacuum black holes}, no. 321, 2008, G\'eom\'etrie diff\'erentielle, physique
  math\'ematique, math\'ematiques et soci\'et\'e. I, pp.~195--265. \MR{2521649}

\bibitem{EHLS}
Michael Eichmair, Lan-Hsuan Huang, Dan~A. Lee, and Richard Schoen, \emph{The
  spacetime positive mass theorem in dimensions less than eight}, J. Eur. Math.
  Soc. (JEMS) \textbf{18} (2016), no.~1, 83--121. \MR{3438380}

\bibitem{Gnullmp}
Gregory~J. Galloway, \emph{Maximum principles for null hypersurfaces and null
  splitting theorems}, Ann. Henri Poincar\'{e} \textbf{1} (2000), no.~3,
  543--567. \MR{1777311}

\bibitem{MOTS06}
\bysame, \emph{Rigidity of marginally trapped surfaces and the topology of
  black holes}, Comm. Anal. Geom. \textbf{16} (2008), no.~1, 217--229.
  \MR{2411473}

\bibitem{GallowayLee}
Gregory~J. Galloway and Dan~A. Lee, \emph{A note on the positive mass theorem
  with boundary}, Lett. Math. Phys. \textbf{111} (2021), no.~4, Paper No. 111,
  10. \MR{4300252}

\bibitem{GalMen2018}
Gregory~J. Galloway and Abra\~{a}o Mendes, \emph{Rigidity of marginally outer
  trapped 2-spheres}, Comm. Anal. Geom. \textbf{26} (2018), no.~1, 63--83.
  \MR{3761653}

\bibitem{GalMen2024}
Gregory~J. Galloway and Abra{\~a}o Mendes, \emph{Some rigidity results for
  compact initial data sets}, Trans. Amer. Math. Soc. \textbf{377} (2024),
  no.~3, 1989–2007.

\bibitem{GS}
Gregory~J. Galloway and Richard Schoen, \emph{A generalization of {H}awking's
  black hole topology theorem to higher dimensions}, Comm. Math. Phys.
  \textbf{266} (2006), no.~2, 571--576. \MR{2238889}

\bibitem{GalSen}
Gregory~J. Galloway and Jos\'{e} M.~M. Senovilla, \emph{Singularity theorems
  based on trapped submanifolds of arbitrary co-dimension}, Classical Quantum
  Gravity \textbf{27} (2010), no.~15, 152002, 10. \MR{2659235}

\bibitem{HE}
Stephen~W. Hawking and George F.~R. Ellis, \emph{The large scale structure of
  space-time}, Cambridge Monographs on Mathematical Physics, vol. No. 1,
  Cambridge University Press, London-New York, 1973. \MR{424186}

\bibitem{Hay}
Sean~A. Hayward, Tetsuya Shiromizu, and Ken-ichi Nakao, \emph{A cosmological
  constant limits the size of black holes}, Phys. Rev. D (3) \textbf{49}
  (1994), no.~10, 5080--5085. \MR{1280410}

\bibitem{DLee}
Dan~A. Lee, \emph{Geometric relativity}, Graduate Studies in Mathematics, vol.
  201, American Mathematical Society, Providence, RI, 2019. \MR{3970261}

\bibitem{LeeLesourdUnger}
Dan~A. Lee, Martin Lesourd, and Ryan Unger, \emph{Density and positive mass
  theorems for initial data sets with boundary}, Comm. Math. Phys. \textbf{395}
  (2022), no.~2, 643--677. \MR{4487523}

\bibitem{LLU23}
\bysame, \emph{Density and positive mass theorems for incomplete manifolds},
  Calc. Var. Partial Differential Equations \textbf{62} (2023), no.~7, Paper
  No. 194, 23. \MR{4612768}

\bibitem{Mars}
Marc Mars, \emph{Stability of marginally outer trapped surfaces and
  applications}, Recent trends in {L}orentzian geometry, Springer Proc. Math.
  Stat., vol.~26, Springer, New York, 2013, pp.~111--138. \MR{3064798}

\bibitem{Nielsen}
Alex~B. Nielsen, \emph{The spatial relation between the event horizon and
  trapping horizon}, Classical Quantum Gravity \textbf{27} (2010), no.~24,
  245016, 14. \MR{2739972}

\bibitem{SYpmt2}
Richard Schoen and Shing-Tung Yau, \emph{Proof of the positive mass theorem.
  {II}}, Comm. Math. Phys. \textbf{79} (1981), no.~2, 231--260. \MR{612249}

\bibitem{Wald}
Robert~M. Wald, \emph{General relativity}, University of Chicago Press,
  Chicago, IL, 1984. \MR{757180}

\end{thebibliography}

\end{document}